\RequirePackage{fix-cm}
\documentclass{svjour3}                     
\smartqed  
\usepackage{graphicx}
\usepackage{bm}
\usepackage{enumitem}
\usepackage[framemethod=tikz]{mdframed}
\newcounter{finding}

\newcommand{\tabincell}[2]{\begin{tabular}{@{}#1@{}}#2\end{tabular}}
\begin{document}

\title{Unveiling Overlooked Performance Variance in Serverless Computing
}


\author{Jinfeng Wen \and Zhenpeng Chen \and Federica Sarro \and Shangguang Wang}


\institute{Jinfeng Wen 
\at Beijing University of Posts and Telecommunications, Beijing, China\\
Beiyou Shenzhen Institute, Beijing, China \\
\email{jinfeng.wen@bupt.edu.cn}
\and
Zhenpeng Chen (corresponding author)
\at Nanyang Technological University, Singapore, Singapore\\
\email{zhenpeng.chen@ntu.edu.sg}
\and
Federica Sarro
\at University College London, London, United Kingdom\\\email{f.sarro@ucl.ac.uk}
\and 
Shangguang Wang
\at Beiyou Shenzhen Institute, Beijing, China \\
Beijing University of Posts and Telecommunications, Beijing, China \\
\email{sgwang@bupt.edu.cn}
}



            

\date{Received: date / Accepted: date}

\maketitle

\begin{abstract}
Serverless computing is an emerging cloud computing paradigm for developing applications at the function level, known as \textit{serverless functions}. 
Due to the highly dynamic execution environment, multiple identical runs of the same serverless function can yield different performance, specifically in terms of end-to-end response latency. However, surprisingly, our analysis of serverless computing-related papers published in top-tier conferences highlights that the research community lacks awareness of the performance variance problem, with only 38.38\% of these papers employing multiple runs for quantifying it.
To further investigate, we analyze the performance of 72 serverless functions collected from these papers. Our findings reveal that the performance of these serverless functions can differ by up to 338.76\% (44.28\% on average) across different runs. Moreover, 61.11\% of these functions produce unreliable performance results, with a low number of repetitions commonly employed in the serverless computing literature. 
Our study highlights a lack of awareness in the serverless computing community regarding the well-known performance variance problem in software engineering. The empirical results illustrate the substantial magnitude of this variance, emphasizing that ignoring the variance can affect research reproducibility and result reliability.
\end{abstract}

\section{Introduction}\label{sec:introduction}

Serverless computing is a rapidly growing cloud computing paradigm. It relieves software developers from the complexities and potential errors of cloud infrastructure management and has been widely adopted in various software applications, such as video processing~\cite{fouladi2017encoding-18}, machine learning~\cite{jiang2021towards-22}, and big data analytics~\cite{muller2020lambada-23}.
Predictions indicate that by 2025, serverless computing will be adopted
by around 50\% of global enterprises~\cite{predictionserverless}. 
Its market size is also expected to skyrocket from \$3 billion in 2017 to \$22 billion by 2025~\cite{marketreport}.

By allowing software developers to focus on application logic, serverless computing empowers them to develop software applications as event-driven, stateless functions, commonly referred to as \emph{serverless functions}.
Leading cloud service providers have introduced dedicated serverless platforms to facilitate the execution of serverless functions, including AWS Lambda~\cite{aws} and Google Cloud Functions~\cite{google}.

The growing presence of serverless computing in software domains has captured the attention of the Software Engineering (SE) field \cite{wen2022literature}. The SE community has studied a broad range of topics about serverless computing, including its development characteristics~\cite{eismann2021state}, programming frameworks~\cite{bermbach2022auctionwhisk}, application migration~\cite{ristov2020daf}, economic impact~\cite{adzic2017serverless}, testing/debugging~\cite{lenarduzzi2020serverless}, and performance optimization~\cite{wen2023FaaSlight}. Notably, performance has emerged as the most widely studied topic within the serverless computing literature~\cite{wen2022literature}. Ensuring accurate and reliable serverless computing performance is critical for research reproducibility and result reliability.

In the SE community, it is well recognized that multiple identical runs of the same application can exhibit varying performance~\cite{pham2020problems,qian2021my,weber2021white,he2019statistics,he2021performance}. Serverless computing-based applications should not be an exception to the \emph{performance variance} phenomenon. Multiple identical runs of the same serverless function can yield different performance, specifically in end-to-end response latency. Several factors can contribute to this performance variance: \textit{1) Highly dynamic cloud underlying infrastructure.} Serverless platforms operate in a cloud environment susceptible to multi-tenancy and networking challenges~\cite{papadopoulos2019methodological,mahgoub2022orion-36}. Moreover, due to opaque instance scheduling and unpredictable invocations of serverless platforms~\cite{patterson2022hivemind-11,fuerst2022locality-32,perron2020starling-26}, the execution environment of serverless functions may be highly variable and dynamic. \textit{2) High-density deployment of lightweight functions.} Serverless functions generally have small memory requirements and are hosted in small instances, leading to high-density deployments~\cite{ShahradATC20,singhvi2021atoll}. This deployment environment increases the risk of disruptions~\cite{zhao2021understanding-34,patterson2022hivemind-11}. 

To date, however, serverless computing research seems to have overlooked the well-known issue of performance variance. To address this gap, in this paper, we present an empirical study that sheds light on this oversight with solid scientific evidence. Additionally, we provide a comprehensive characterization of the magnitude of performance variance in serverless computing, emphasizing the significant consequences of disregarding this crucial aspect.

To this end, we collect and analyze 99 research papers related to serverless computing performance from 77 top-tier academic conferences. Our findings reveal that only 38.38\% of these research papers employ multiple runs to quantify the variance in serverless function performance. 
This shows that the current serverless computing literature lacks awareness for the well-known performance variance problems in SE.

Furthermore, we extract 72 serverless functions from these research papers and measure their end-to-end response latencies across multiple runs to assess the extent of performance variance. Our analysis demonstrates that serverless function performance can vary significantly, with a maximum variance of 338.76\% (44.28\% on average) observed among different runs. 

Finally, we assess the reliability of serverless function performance obtained at commonly used numbers of repetitions within the serverless computing literature. Our investigation uncovers that under these conditions, 
98.61\% of the serverless functions require at least 50 executions to achieve reliability. 
This emphasizes the need for significantly more repetitions in measuring the serverless function performance, surpassing conventional practices in the existing literature.

Our findings offer practical implications for different stakeholders in the research community. \textbf{1)~Researchers} should thoroughly assess serverless function performance variance, detailing their measurement approach, including repetitions and reasoning, to ensure research reproducibility and result reliability. Furthermore, a potential research space needs to be explored to develop novel performance testing techniques specialized for serverless computing, enhancing the accuracy and reliability of performance assessments.
\textbf{2)~Software developers} can customize strategies to mitigate the severe performance variance in serverless computing-based applications, ensuring consistent user experiences. \textbf{3)~Cloud providers} offering serverless computing environments can provide consistent quality of service when delivering the services in a serverless computing manner to meet the demands of customers who seek stable performance.

In summary, this paper makes the following contributions:
\begin{itemize}[leftmargin=*]
\item An empirical study highlighting the lack of awareness within the serverless computing community regarding the well-known performance variance problem in software engineering.
\item A measurement study that illustrates the substantial magnitude of performance variance in serverless computing, underscoring the significant consequences of neglecting this critical aspect.
\item A replication package~\cite{ourdata} including the paper's data and scripts, serving as a benchmark dataset for future research on performance in serverless computing.
\end{itemize}

\section{Background}\label{sec:background}
We start by introducing the background knowledge of serverless computing and performance of serverless functions.


\subsection{Serverless computing}


Serverless computing is a popular cloud computing paradigm, which allows software developers to focus on their application logic without having to manage complex cloud underlying tasks. Function-as-a-Service (FaaS) is the most prominent implementation of serverless computing~\cite{AkkusATC18,taibi2020serverless,WenServerless21,copik2021sebs,JonasCoRR2019new}. Thus, in this study, we focus on FaaS.
In the FaaS fashion, software developers implement applications as a series of event-driven and stateless dedicated functions, called \textit{serverless functions}. They deploy and execute serverless functions on serverless platforms, such as AWS Lambda~\cite{aws} and Google Cloud Functions~\cite{google}.


\begin{figure}[t]
	\centering
    \includegraphics[width=0.7\textwidth]{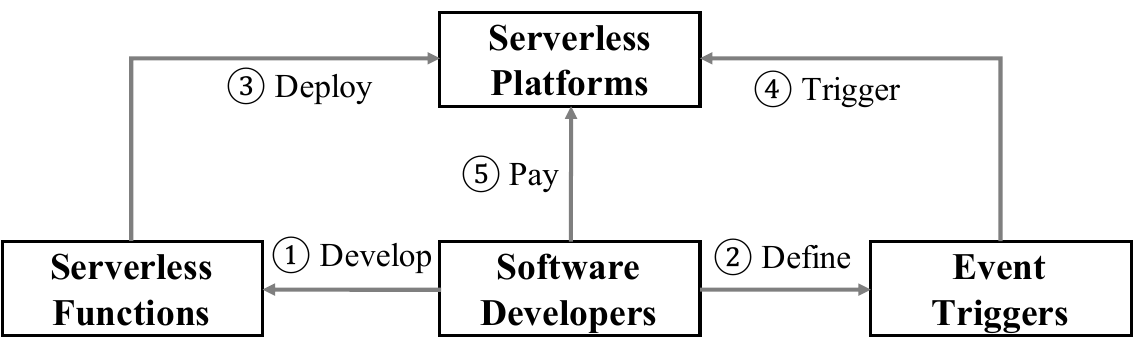}
    \caption{The process of developing, deploying, and executing serverless functions by software developers.
    }
    \label{fig:developmentprocess}
\end{figure}


Fig.~\ref{fig:developmentprocess} illustrates the process of developing, deploying, and executing serverless functions. \textcircled{1} First, software developers implement serverless functions using programming interfaces as event-driven and stateless functions~\cite{wen2022literature}. These functions are generally written in high-level programming languages, e.g., Python and JavaScript~\cite{serverlesscommunitysurvey,serverlesscommunitysurvey1,eskandani2021wonderless}. \textcircled{2} Meanwhile, software developers can define the rules to bind their serverless functions and the related events, such as HTTP requests and data changes of cloud storage. \textcircled{3} Then, serverless functions and required dependent libraries are packaged together and deployed to serverless platforms. In this stage, software developers can specify the required configurations, including language runtime, memory size, and function timeout time~\cite{wen2022literature}. 
\textcircled{4} When the serverless functions are triggered by pre-defined events, the serverless platforms can automatically launch or reuse the required function instances (e.g., VMs or containers) with restricted resources (e.g., CPU and memory) to serve these requests. This frees software developers from complex server management and makes their requests benefit from seamless resource elasticity. \textcircled{5} After the requests are completed, software developers pay for the number of requests and the actually allocated or consumed resources~\cite{lin2020modeling,wen2022literature}.




\subsection{Performance of serverless functions} 
In this study, we focus on evaluating the performance of serverless functions by examining the end-to-end response latency (abbreviated as \textit{e2eTime}). The end-to-end response latency is a widely adopted metric in the serverless computing literature~\cite{yu2022accelerating,scheuner2020function,lin2020modeling,wen2023FaaSlight,wen2021characterizing}. It is defined as the duration as when a request is sent from a client until its completion.
This metric is considered the primary metric for assessing serverless function performance, and serverless computing research actively proposes optimization techniques targeting this metric~\cite{wang2021faasnet-53,mahgoub2022orion-36,zhao2021understanding-34}. Additionally, practitioners commonly use end-to-end response latency as a reflection of user experience satisfaction~\cite{eismann2021state,wen2021characterizing,WenServerless21}.

Generally, the response latency of a serverless function is divided into cold-start response latency and warm-start response latency. (1) When a serverless platform launches new function instances to handle requests, the serverless function encounters cold-start invocations. This kind of invocation involves the preparation process of function instances.
Thus, this latency is referred to as the cold-start response latency. (2) On subsequent invocations of the same serverless function within a short keep-alive time, the serverless platform can reuse the previously launched function instances. This leads to warm-start invocations for the serverless function and produces warm-start response latency. In the absence of requests, the serverless platform automatically turns the launched function instances into an idle state and releases the corresponding resources. Generally, warm-start response latency tends to be smaller than the cold-start response latency for the same serverless function due to the absence of the preparation process of function instances during warm starts.

\section{Methodology}\label{sec:method}

\begin{figure}[t]
	\centering
    \includegraphics[width=0.7\textwidth]{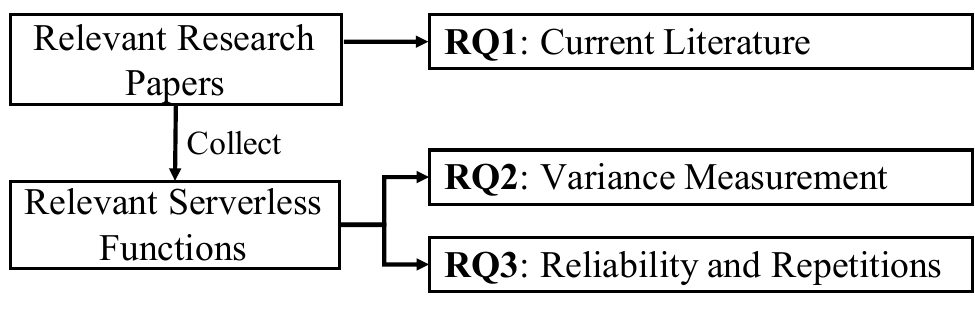}
    \caption{An overview of our methodology.}
    \label{fig:overview}
\end{figure}

To understand the variance of serverless function performance, we first identify relevant research papers and collect serverless functions from them. Based on them, we answer three research questions (RQs) related to the variance of serverless function performance. In Fig.~\ref{fig:overview}, we show an overview of our methodology.

\subsection{Research questions}

\textbf{RQ1 (Current literature):} \textit{To what extent has the current literature addressed the variance of serverless function performance?}
This RQ investigates whether and how researchers consider the performance variance of serverless functions when studying serverless computing, given that performance variance has been a well-known issue in SE.


\noindent \textbf{RQ2 (Variance measurement):} \textit{How variable is the performance of serverless functions?}
Despite the well-known performance variance issue, the magnitude of this variance in serverless computing remains unclear. 
If the variance is substantial, researchers and developers cannot disregard it during the development of serverless computing-based applications.
This RQ aims to fill this knowledge gap and provide actionable insights for stakeholders.

\noindent \textbf{RQ3 (Reliability and repetitions):} \textit{How reliable are the serverless function performance results obtained from different repetitions?} This RQ aims to explore the reliability of serverless function performance across multiple repetitions, highlighting the importance of repetition settings on the reliability of performance results. Additionally, it seeks to provide guidance on determining the appropriate number of repetitions needed to achieve more reliable performance measurements.







\subsection{Collection of relevant papers}

\begin{figure}[t]
	\centering
    \includegraphics[width=0.7\textwidth]{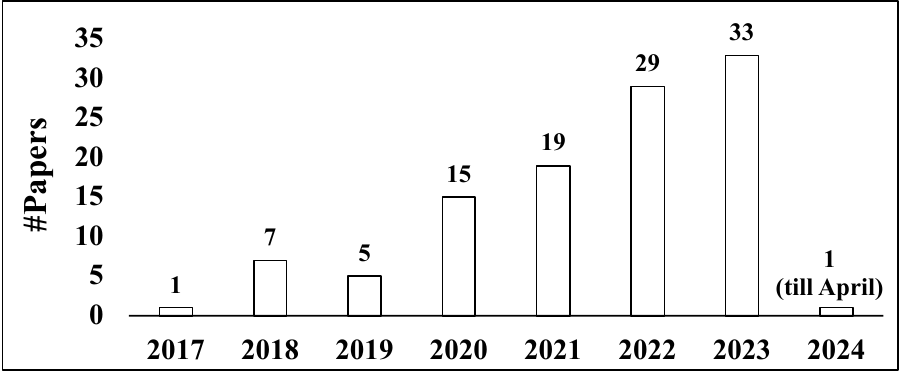}
    \caption{Number of serverless computing-related papers in top-tier conferences per year.}
    \label{fig:PaperDistribution}
\end{figure}

\begin{figure}[t]
	\centering
    \includegraphics[width=0.8\textwidth]{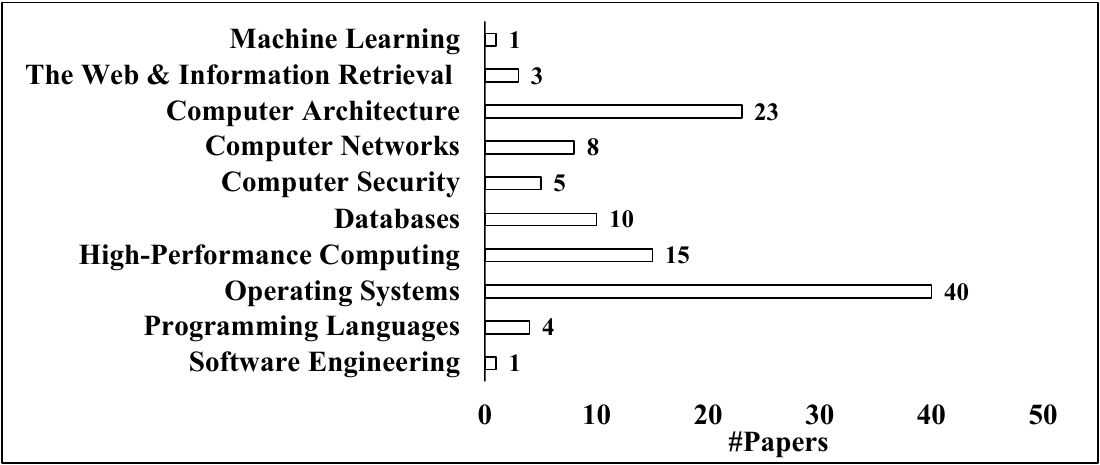}
    \caption{Distribution of 110 serverless computing-related papers per computer science area.}
    \label{fig:AreaDistribution}
\end{figure}

To address RQ1, we collect research papers related to serverless computing. Since studies on serverless computing have been published in academic venues across various research communities~\cite{wen2022literature}, we gather research papers from all 77 conferences listed in the Computer Science Rankings (CSRankings)~\cite{csrankings}. CSRankings provides a curated list of top-tier conferences from various areas of computer science, including Software Engineering (e.g., ICSE, FSE, ASE, and ISSTA), Operating Systems (e.g., OSDI, SOSP, EuroSys, FAST, and ATC), and Computer Networks (e.g., SIGCOMM and NSDI). These conferences are recognized not only for their academic reputation but also for their influence in the respective research communities. We focus on papers published in these high-tier conferences because they reflect state-of-the-art research and provide a comprehensive view of the latest findings in serverless computing across diverse areas.
We do not use search engines (e.g., Google Scholar) and citation-based thresholds to collect impactful papers related to serverless computing. This is because defining a consistent and objective citation threshold is challenging, as citation counts can be influenced by factors such as publication date, research topic popularity, and self-citation. Moreover, there is no universally accepted standard for what constitutes ``impactful'' based on citations. We do not evaluate the quality of papers based on their authors or assume that work published in other venues is of lower quality, as such judgments could introduce bias. Instead, we adopt a venue-based selection criterion, relying on CSRankings to identify high-quality, peer-reviewed research from top-tier conferences across various areas of computer science. This approach minimizes subjectivity and ensures consistency.
However, we acknowledge that including papers from other venues (such as journals and other conferences) in our study would provide a more comprehensive view of the research landscape. In future work, we plan to expand our study to include more research papers to further enhance the scope of our research.


The process for collecting research papers is as follows.
First, we collect all technical papers (excluding short papers, tutorials, posters, workshop papers, industry track papers, etc.) published in the 77 conferences between 2014 (the year serverless computing was popularized~\cite{WenServerless21,wen2022literature,JonasCoRR2019new}) and the date we collect the papers (April 10, 2024).
Then, the first two authors independently review the research papers to select those related to serverless computing. A paper is considered relevant if, upon manual inspection of its title, abstract, and full text, the authors determine that it discusses or addresses specific problems in the field of serverless computing.
If the authors disagree on the relevance of a paper, an arbitrator, who has ten years of cloud computing experience, is involved in discussing to reach an agreement. To measure the inter-rater agreement level of the authors during the independent labeling, we use Cohen's Kappa ($\kappa$)~\cite{cohen1960coefficient}, which is the most widely-used agreement evaluation metric in SE~\cite{uta2020big,WenServerless21,wen2022literature}. The value of $\kappa$ is 0.952, indicating an almost perfect agreement and a reliable labeling procedure~\cite{landis1977measurement}.
Finally, we obtain 110 research papers related to serverless computing. 
The number of related papers per year is shown in Fig.~\ref{fig:PaperDistribution}. 
We observe an overall increasing trend, with the number of relevant papers rising from 1 in 2017 to 33 in 2023. This implies that serverless computing is gaining increasing attention from the research community, demonstrating the timeliness and urgency of our study.
The number of serverless computing-related papers collected from various areas of computer science is illustrated in Fig.~\ref{fig:AreaDistribution}. The three leading areas are Operating Systems, Computer Architecture, and High-Performance Computing, which collectively account for 40, 23, and 15 papers, respectively.
Furthermore, the first two authors jointly filter papers that do not report the performance results of serverless functions. 
As a result, 99 out of the 110 research papers are retained for our final performance analysis. 
This is consistent with findings reported by previous work~\cite{wen2022literature,li2023serverless,scheuner2020function} that most serverless computing papers are related to serverless function performance. 
Specifically, the distribution is as follows: 5 papers in 2019 are reduced to 4 papers, 15 papers in 2020 are reduced to 14, 19 papers in 2021 are reduced to 15, 29 papers in 2022 are reduced to 26, and 33 papers in 2023 are reduced to 31.
Details of the selected papers (including conference name, year, and paper title) are provided in our repository~\cite{ourdata}.




\subsection{Collection and analysis of serverless functions}

To answer RQ2 and RQ3, we extract serverless functions used in the 99 research papers. 
We follow previous work~\cite{wen2023FaaSlight} to focus on serverless functions developed using the most widely adopted \textit{serverless platforms} (i.e., AWS Lambda and Google Cloud Functions) and \textit{programming languages} (i.e., Python and JavaScript).
Moreover, we select serverless functions that are open-sourced and have the corresponding guidance documentation to assist with the execution. 
According to these criteria, the first two authors jointly extract and finally collect 72 serverless functions covering a wide range of tasks, such as Web request handling, video processing, scientific computing, machine learning, and natural language processing. Most of the collected serverless functions are included in commonly used benchmarks in serverless computing research~\cite{yu2020characterizing,kim2019functionbench,maissen2020faasdom} and industry~\cite{awsserverlessrepo}, such as \textit{FunctionBench}~\cite{kim2019functionbench}, \textit{ServerlessBench}~\cite{yu2020characterizing}, \textit{AWS Sample}~\cite{AWSSamples}, \textit{SeBS}~\cite{copik2021sebs}, and \textit{FaaSDom}~\cite{maissen2020faasdom}.
Among these serverless functions, 67 serverless functions are executed on AWS Lambda, while 5 serverless functions are executed on Google Cloud Functions. It can be because the advent of AWS Lambda was the main reason for the popularity of serverless computing~\cite{wen2022literature,JonasCoRR2019new,WenServerless21}. Moreover, 59 serverless functions are written in Python, while 13 serverless functions are written in JavaScript. Such a distribution can be attributed to the increasing number of tasks that use popular Python third-party libraries~\cite{wen2022literature}. We denote the 72 serverless functions as \textit{Func1}, \textit{Func2}, ..., and \textit{Func72}. The number mapping information about serverless functions is provided in our repository~\cite{ourdata}. 




We then analyze the performance of the 72 serverless functions. We execute them with the input, configurations, and serverless platform used in their original papers. If no specific input is given in the original paper, we construct an input that matches the task execution according to the functionality of the serverless function.
If no specific configurations are given, serverless functions are executed using the default configurations of serverless platforms, as developers typically rely on default configurations.
At the time of our study, the default memory size of AWS Lambda is 128 MB~\cite{AWSmemory}, and its function timeout time is 3 seconds~\cite{AWStimeout}. 
For Google Cloud Functions, the default memory size and function timeout time are 256 MB~\cite{Googlememory} and 60 seconds~\cite{Goolgetimeout}, respectively.
We execute serverless functions to detect the sufficiency of their memory size or function timeout time. If the memory size or timeout is found to be insufficient during execution, the serverless platform reports errors and indicates the actual memory used or that the timeout is nearing its limit. In such cases, we continue to execute the functions by gradually increasing the memory size or function timeout until successful execution is achieved.
To minimize the potential impact of the network variability on our measurements, we adopt the following strategies: (i) We use the same experimental machine in the same geographical area to send requests to the corresponding serverless platform and receive responses. (ii) We deploy our measured serverless functions to the same service region across different serverless platforms, such as ``us-west-1'' of AWS Lambda and ``us-west1'' of Google Cloud Functions. 

To explore performance results across different times of the day and capture the comprehensive performance characteristics of serverless functions, we obtain response latencies of serverless functions for 50 trials. Meanwhile, to explore different performance types (especially ``cold-start performance'') of serverless functions, we employ a different strategy of multiple interleaved trials, meaning that functions from different platforms and languages are executed in an interleaved manner, with each trial being separated by a fixed interval (e.g., half an hour). This strategy also mitigates the issue of functions being executed too closely in time and makes all performance results span at least one full day to capture variability across different times of the day.
In each trial, we execute each tested serverless function twice using the same input and configuration to obtain the cold-start response latency and warm-start response latency, respectively.
We set the fixed interval between each trial to half an hour, because we find that the resources used for execution can be released to ensure that the start of the next trial is still a cold start. This also reduces potential residual effects on subsequent trials. In addition, in each trial, the second execution of each function starts five seconds after the completion of the first execution, because we find that this duration can ensure the successful warm start of each serverless function.
Finally, we obtain 72 $\times$ (50 + 50) = 7,200 data points (3,600 for cold start and 3,600 for warm start) about the end-to-end response latency of serverless functions. 
The execution results are produced between April 17, 2024 and May 10, 2024.

\section{RQ1: Current Literature}\label{sec:literature}


RQ1 investigates whether and how the performance variance of serverless functions has been considered in the serverless computing literature, as this affects the reproducibility of conclusions drawn in research. To this end, we follow the standard criteria established by previous work~\cite{uta2020big} to analyze our selected 99 research papers related to serverless computing. Specifically, Uta et al.~\cite{uta2020big} defined standard criteria to determine if a study’s conclusions are reproducible under performance variability, including checking whether:

\begin{itemize}[leftmargin=*]
 \item (1) Mean or median performance values are reported;
 \item (2) Confidence (e.g., confidence intervals) or variability (e.g., standard deviation, coefficient of variance, or percentiles) is reported;
 \item (3) The number of times an experiment was repeated is reported.
\end{itemize}



Such information can assist us in further understanding the reliability of the provided evidence and the study's conclusions. The first two authors separately read the full text of each paper to find out if it reports the aforementioned information. 
Moreover, if a paper reports the experiment repetition, i.e., \textit{criterion (3)}, we also record the specific number of repetitions used in the paper. 
A paper may involve multiple pieces of information, which may result in the percentage of the total number of papers exceeding 100\% when the information is counted.
The inter-rater agreement values of labeling, measured by Cohen's Kappa ($\kappa$)~\cite{cohen1960coefficient}, are 0.927, 0.914, and 0.893 for each type of information, respectively, indicating perfect agreement and a reliable labeling process~\cite{landis1977measurement}. During the labeling process, encountered conflicts are discussed to reach an agreement by the first two authors and the third arbitrator.


\textbf{Results:} We observe that researchers lack awareness of the performance variance of serverless functions and often do not adequately describe related experimental settings, despite the fact that performance variance has become a well-known problem in SE. 
Specifically, Table.~\ref{tab:PaperMetric} shows the results of our literature analysis. We observe that 62.63\% (62 out of 99) of the papers report means or medians; only 37.37\% (37 out of 99) report statistical variation data (e.g., percentiles and confidence intervals); and only 38.38\% (38 out of 99) use multiple runs when conducting serverless computing-related experiments or performance analyses.
In other words, over 60\% of the research papers do not report how many times they repeated the experiments. This omission suggests that either no repetitions were performed or a specific number of repetitions was used but not disclosed. This lack of information affects the reproducibility of the results. 
We examine the details on reporting experiment repetitions in 99 final papers from various areas of computer science, shown in Table~\ref{tab:rq1repetitiondetails}. The values are formatted as $a/b$, where $a$ represents the number of papers that use experiment repetitions in each specific area and $b$ is the total number of papers in this area. We observe that The Web\&Information Retrieval has the largest representation with 36 papers, of which 15 papers report experiment repetitions. The areas of Computer Architecture (15 total papers) and High-Performance Computing (22 total papers) also show notable representation, with 7 and 6 papers, respectively, reporting experiment repetitions. Other areas, such as Machine Learning (2/3), Computer Networks (3/8), and Databases (1/7), have smaller numbers of papers reporting repetitions. Certain areas, e.g., Programming Languages, have minimal representation, with only 1 paper, which does not report repetitions (0/1). In summary, the reporting of experiment repetitions varies across the areas of computer science.


\begin{table*}[t]
\footnotesize
 \caption{(RQ1) The details about experiment reporting for the studied 99 research papers.}
    \label{tab:PaperMetric}
    \begin{tabular}{c|c|c|c}
    \hline
        & \tabincell{c}{\textbf{Reporting Mean} \\ \textbf{or Median}} & \tabincell{c}{\textbf{Reporting Statistical} \\ \textbf{Variation Data}} & \tabincell{c}{\textbf{Reporting the} \\ \textbf{Number of Repetitions}}   \\
    \hline
     \tabincell{c}{\textbf{\#Papers}} & 62.63\% & 37.37\% & 38.38\%  \\ \hline

\end{tabular}
\end{table*}





\begin{table*}[t]
\footnotesize
 \caption{(RQ1) Details about experiment repetitions reported in 99 final research papers from various areas of computer science}, formatted as $a/b$. $a$ represents the number of papers that use experiment repetitions in each specific area and $b$ is the total number of papers in this area.
    \label{tab:rq1repetitiondetails}
    \begin{tabular}{c|c|c|c|c}
    \hline
      \tabincell{c}{Machine \\ Learning} & \tabincell{c}{The Web\&Information \\ Retrieval} & \tabincell{c}{Computer \\ Architecture} & \tabincell{c}{Computer \\Networks} &  \tabincell{c}{Computer \\Security}  \\
    \hline
     2/3 & 15/36 & 7/15 & 3/8 & 2/4 \\ \hline
      \hline
     \tabincell{c}{Databases} & \tabincell{c}{High-Performance \\ Computing} & \tabincell{c}{Operating \\Systems} & \tabincell{c}{Programming \\ Languages} &  \\
    \hline
     1/7 & 6/22 & 1/3 & 0/1 & \\ \hline

\end{tabular}
\end{table*}

\begin{figure}[t]
	\centering
    \includegraphics[width=0.7\textwidth]{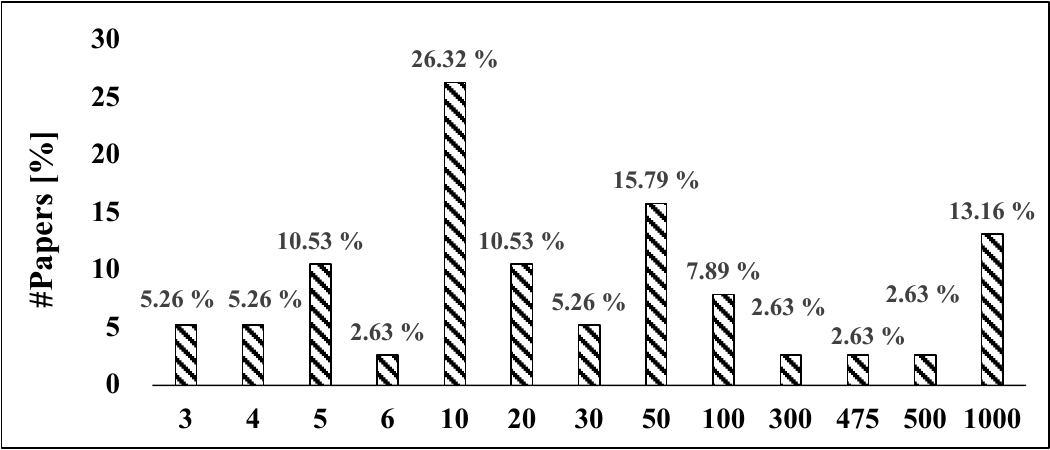}
    \caption{(RQ1) Distribution of the number of repetitions used in the research papers that report experiment repetitions.}
    \label{fig:PaperRepetition}
\end{figure}

We further analyze the papers that report experiment repetitions.
Fig.~\ref{fig:PaperRepetition} shows the number of repetitions used in these papers. Note that a paper can employ multiple numbers of repetitions, so the sum of the proportions in the figure may exceed 100\%.
There are 13 kinds of numbers of repetitions in the serverless computing literature. We find that 81.58\% of these papers use no more than 50 repetitions.
The top 3 frequently used repetitions are 10, 50, and 1,000, which account for 26.32\%, 15.79\%, and 13.16\% of these papers, respectively. 

Surprisingly, we find that none of these papers justify or explain the reason for the selection of the repetition number. In fact, the selection of this number is of high significance.
If the number of repetitions is too low (e.g., 10 repetitions), the reliability of the obtained performance of serverless functions may not be guaranteed (as explained in Section~\ref{sec:credible}), and it could lead to wrong or ambiguous conclusions. If the number of repetitions is too large (e.g., 1,000 repetitions), it may cause a huge runtime overhead, e.g., long experimentation time and high costs. In serverless computing, researchers need to pay for the number of requests and the actually allocated or consumed resources of serverless functions~\cite{lin2020modeling,wen2022literature}. In this situation, a large repetition number can produce a huge cost. 
Thus, a too-small or too-large number of repetitions may be inappropriate in serverless computing-related experiments. A careful selection of the number of repetitions is needed in these experiments.



\vspace{2.3mm}


\begin{mdframed}[linecolor=gray,roundcorner=12pt,backgroundcolor=gray!15,linewidth=3pt,innerleftmargin=2pt, leftmargin=0cm,rightmargin=0cm,topline=false,bottomline=false,rightline = false]
    \textbf{Finding 1}: 
    The current serverless computing literature lacks awareness of the well-known performance variance problem in software engineering. Specifically, only 38.38\% of the relevant papers use multiple runs to quantify the performance variance of serverless functions. 81.58\% of the papers that report experiment repetitions use no more than 50 repetitions. Additionally, none of the papers provide any relevant description to justify the reason for the number of repetitions that they choose.     
    
\end{mdframed}

\section{RQ2: Variance Measurement}\label{sec:measure}
RQ2 explores the magnitude of the performance variance of serverless functions. To this end, we analyze response latencies of 72 serverless functions over multiple runs from two aspects: \textit{coefficient of variance (CV)} and \textit{boxplot}, which are commonly used in performance variance analysis~\cite{maricq2018taming,guizzo2020cost,zhao2021understanding-34,uta2020big,pham2020problems}. 
Specifically, CV, known as the dispersion coefficient, is a statistical indicator that measures the degree of data variability. It represents the ratio of the standard deviation to the mean. Compared to standard deviation, CV provides a normalized, relative measure of dispersion and allows for comparison between data distributions with different units or means. A big value of CV means a large degree of performance variance.
The boxplot can provide a visual representation of the dispersion degree for the performance results generated by each serverless function over multiple repetitions.

We use response latencies of each function over 50 repetitions to explore the magnitude of the variance. We chose 50 repetitions because this number is not less than the settings in over 80\% of existing papers reporting repeated runs and is actually an upper value among them. We adopt this upper value to assume that most papers offer improved performance compared to the original setup. We investigate whether these assumed improved performance results suffer from a large magnitude of performance variance.



\begin{figure}[t]
	\centering
    \includegraphics[width=0.7\textwidth]{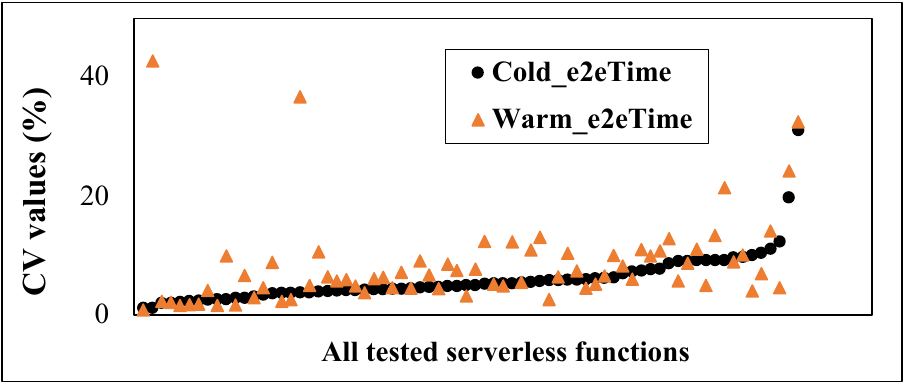}
    \caption{(RQ2) CV values for cold-start response latency are ordered in ascending order. CV values for warm-start response latency are placed according to serverless functions that have been sorted by the cold-start CV values.}
    \label{fig:CVvalues}
\end{figure}



\textbf{Results:} 
(1) \textit{CV:}
We analyze CV values calculated from response latencies generated by each serverless function under cold start and warm start. Fig.~\ref{fig:CVvalues} shows these results, where cold-start CV values are ordered in ascending order. 


First, we observe that response latencies of serverless functions have different degrees of variance under cold and warm starts. Specifically, cold-start CV values range from 1.23\% to 31.18\%, while warm-start CV values range from 0.86\% to 42.81\%, differing by as much as 49 times and thus showing a wide range of variance. On average, the cold-start CV value is 6.07\%, while the warm-start CV value is 8.39\%. The median of cold-start CV values is 5.06\%, while that of warm-start CV values is 6.47\%. These results indicate that functions executed in cold and warm starts produce different degrees of variance in the end-to-end response latency. 

\begin{figure}[t]
	\centering
    \includegraphics[width=0.7\textwidth]{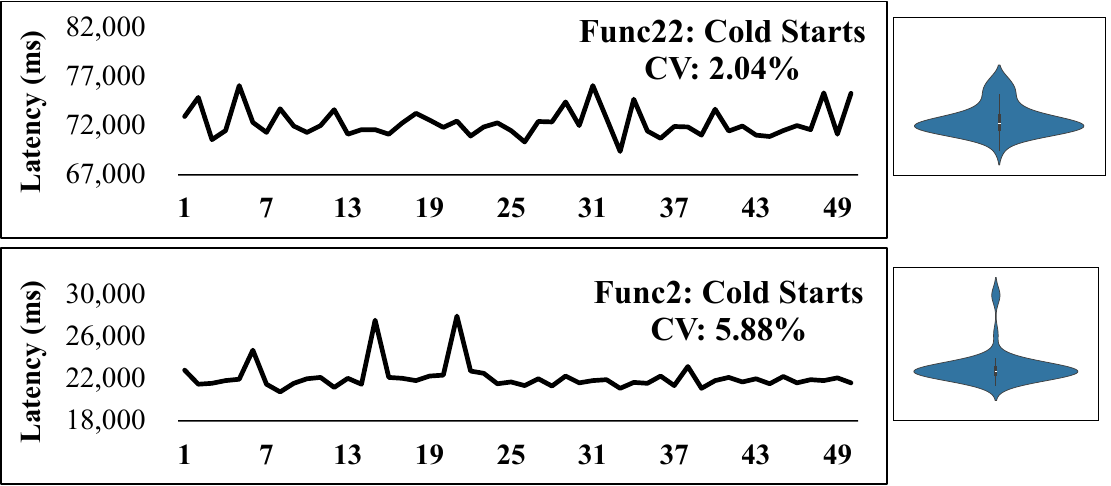}
    \caption{(RQ2) Example of the generated serverless function performance. \normalfont{The left part shows 50 data points obtained in cold starts for \textit{Func22} and \textit{Func2}. The right part is the violin plot, whose thickness is proportional to the probability density of the data.}}
    \label{fig:STDExample}
\end{figure}

However, providing only CV values is not a straightforward way to understand how the performance of serverless functions actually fluctuates. Thus, we give examples of the raw performance results of the serverless function. 
Fig.~\ref{fig:STDExample} shows the performance data (50 data points) from cold starts for two serverless functions, $Func22$ and $Func2$. Multiple runs of the same function produced fluctuating response latencies, with no regular patterns of change. We also check the data from other serverless functions, but no consistent change patterns are found. The changes remain irregular, as seen in the data points in Fig.~\ref{fig:STDExample}.
The CV values of \textit{Func22} and \textit{Func2} are 2.04\% and 5.88\%, respectively. The difference between the maximum and minimum values of \textit{Func22} achieves 6758.63 milliseconds (approximately 7 seconds), and that of \textit{Func2} is 7244.32 milliseconds (also about 7 seconds). Generally, serverless functions execute short-lived and milliseconds-level tasks~\cite{singhvi2021atoll,KlimovicWSTPK18}. Thus, the second-level difference size is severe for serverless functions. Moreover, from the right violin plot of Fig.~\ref{fig:STDExample}, the distributions of data points of performance for different serverless functions are drastically different, being clustered in different positions. The data points of \textit{Func22} are distributed near the middle of its violin plot, while those of \textit{Func2} are distributed towards the bottom. Overall, we observe a significant variance in serverless function performance from two examples, where CV values are 2.04\% and 5.88\%, respectively. Therefore, we infer that other serverless functions with larger CV values than those shown in the examples will produce more significant fluctuations in performance.

We check the top 5 serverless functions with high CV values. For cold or warm starts, four serverless functions are executed on AWS Lambda and one on Google Cloud Functions. 
This observation demonstrates that serverless functions with high CV values are not limited to a single serverless platform. The programming languages used in these functions include Python and JavaScript, which are not limited to a single programming language. 

We further analyze the serverless functions with high CV values, e.g., \textit{Func30} with warm-start CV value of 42.81\%. The maximum and minimum values of response latencies of \textit{Func30} differ by 338.76\%, where the maximum value reaches 4.39 times its minimum value. 
The possible reason is that \textit{Func30} needs to connect the cloud storage (e.g., AWS S3~\cite{awss3}) to transfer the file to the function instance through the network. However, the functionality of serverless functions that rely on platform library code to establish connections and transfer files with cloud services may lack robust design. Moreover, the network may be unstable. These likely result in unpredictable performance fluctuations. We check all serverless functions, and the corresponding maximum and minimum values of the response latency differ by a mean of 42.28\% and a median of 29.67\%.

To further investigate the impact of platforms and languages, we compare the latency differences between platforms and languages, as shown in Table~\ref{tab:rq2comparison}. \textbf{Platform Comparison}: We compare the variability between Google Cloud Functions and AWS Lambda. For the mean, AWS Lambda shows a latency difference of 40.06\%, whereas Google Cloud Functions shows  72.02\%. In terms of median results, AWS Lambda has a latency difference of 29.24\%, while Google Cloud Functions has 47.39\%. This indicates that Google Cloud Functions experiences a more significant performance variability than AWS Lambda. \textbf{Language Comparison}: We compare Python and JavaScript. On average, Python-based functions exhibit a latency difference of 39.49\%, while JavaScript functions show 54.92\%. For the median, Python has a 29.24\% difference, while JavaScript has 31.70\%. This suggests that JavaScript has a higher performance variability than Python. \textbf{Platform and Language Interaction}: We explore the interaction between platforms and languages. Python-based serverless functions executed on AWS Lambda have a latency difference of 39.30\%, whereas those on Google Cloud Functions show 44.85\%. From the results, serverless functions written in JavaScript and executed on Google Cloud Functions exhibit the highest variability in both maximum and minimum latencies.

\begin{table*}[t]
\footnotesize
 \caption{(RQ2) A comparison of the performance variability across serverless platforms and languages by calculating the latency difference of the maximum and minimum values of the serverless function.}
    \label{tab:rq2comparison}
    \begin{tabular}{c|c|c|c|c}
    \hline
        & \tabincell{c}{AWS Lambda} & \tabincell{c}{Google Cloud \\ Functions} & \tabincell{c}{Python} & \tabincell{c}{JavaScript}   \\
    \hline
     \tabincell{c}{Mean} & 40.06\% & 72.02\% & 39.49\% & 54.92\% \\ \hline
      \tabincell{c}{Median} & 29.24\% & 47.39\% & 29.24\% & 31.70\% \\ \hline
      \hline
    & \tabincell{c}{AWS Lambda \\ + Python} & \tabincell{c}{Google Cloud \\ Functions+Python} & \tabincell{c}{AWS Lambda \\+JavaScript} & \tabincell{c}{Google Cloud \\ Functions+JavaScript}   \\
    \hline
     \tabincell{c}{Mean} & 39.30\% & 44.85\% & 44.36\% & 90.14\% \\ \hline
      \tabincell{c}{Median} & 28.76\% & 47.39\% & 31.70\% & 46.20\% \\ \hline

\end{tabular}
\end{table*}

We further investigate how performance variance relates to the characteristics of the tasks by providing representative examples. We categorize some functions into three common types: CPU-memory-intensive tasks, I/O-intensive tasks, and network-intensive tasks. To explore this, we select three serverless functions from each category: for CPU-memory-intensive tasks, we examine $Func5$, $Func6$, and $Func7$; for I/O-intensive tasks, $Func10$, $Func11$, and $Func33$; and for network-intensive tasks, $Func14$, $Func37$, and $Func38$. The latency differences between these task types are summarized in Table~\ref{tab:rq2comparison2}. Specifically, CPU-memory and network-intensive tasks show a latency difference of more than 30\%, while I/O-intensive tasks exhibit a difference exceeding 20\%. Similar trends are observed when considering median latency values. The results reveal different latency differences influenced by task characteristics, which serve as a preliminary exploration of the impact of task characteristics on performance variance.

\begin{table*}[t]
\footnotesize
 \caption{(RQ2) A comparison of performance variance of serverless functions in the context of their exhibited characteristics by calculating the latency difference of the maximum and minimum values.}
    \label{tab:rq2comparison2}
    \begin{tabular}{c|c|c|c}
    \hline
        & \tabincell{c}{CPU-Memory Task} & \tabincell{c}{I/O Task} & \tabincell{c}{Network Task}    \\
    \hline
    \tabincell{c}{Functions} & Func5, Func6, Func7 & Func10, Func11, Func33 & Func14, Func37, Func38  \\ \hline
     \tabincell{c}{Mean} & 38.94\% & 27.06\% & 34.45\%  \\ \hline
      \tabincell{c}{Median} & 34.68\% & 21.77\% & 36.60\% \\ \hline

\end{tabular}
\end{table*}

Then, we observe that the variance of serverless function performance is more severe under warm starts than under cold starts. Specifically, the warm-start CV values for 65.28\% (47/72) of the serverless functions are greater than those of the corresponding cold starts. We also calculate the number of the functions whose CV values are greater than 10\%, which generally indicates a large degree of variance~\cite{maricq2018taming,zhao2021understanding-34,uta2020big}. In cold starts, there are 8.33\% (6/72) functions, while warm starts have 29.17\% (21/72). This implies that warm-start response latency has a more severe variance than cold-start response latency. One possible reason is that executing tasks in the reused function instances (i.e., warm starts) may be more susceptible to resource contention and underlying policies in serverless platforms. Another possible reason is that warm-start response latency tends to be smaller than cold-start response latency. The similar latency difference sizes may create the illusion of higher (lower) CV values when observed at smaller (larger) values. 

\begin{figure*}[t]
	\centering
    \includegraphics[width=0.99\textwidth]{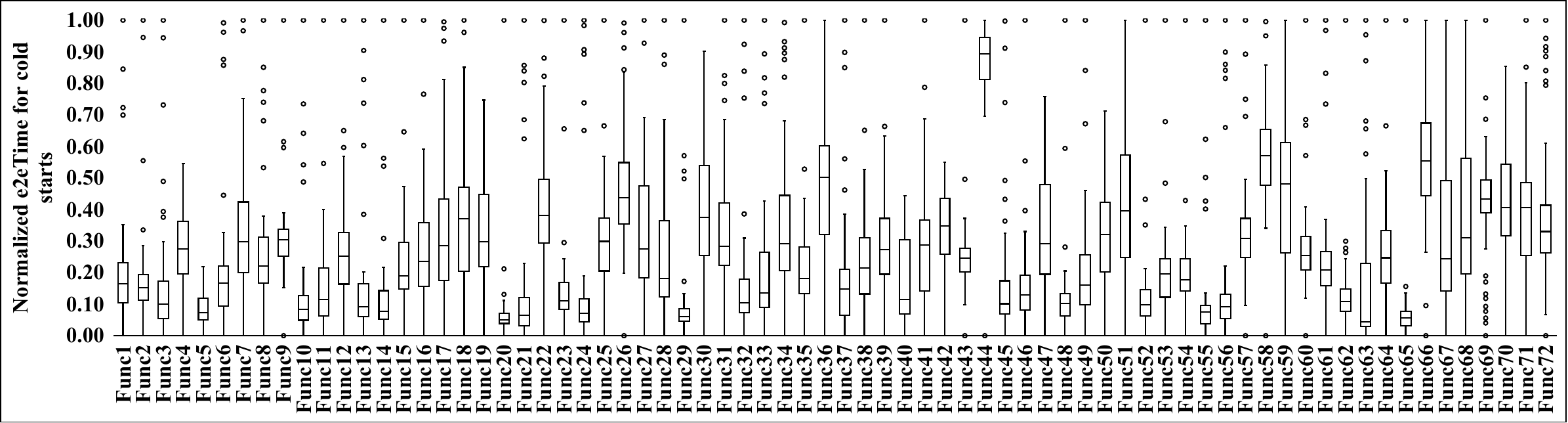}
    \caption{(RQ2) The normalized boxplot for the cold-start response latencies of 72 serverless functions.}
    \label{fig:boxplotcold}
\end{figure*}
\begin{figure*}[t]
	\centering
    \includegraphics[width=0.99\textwidth]{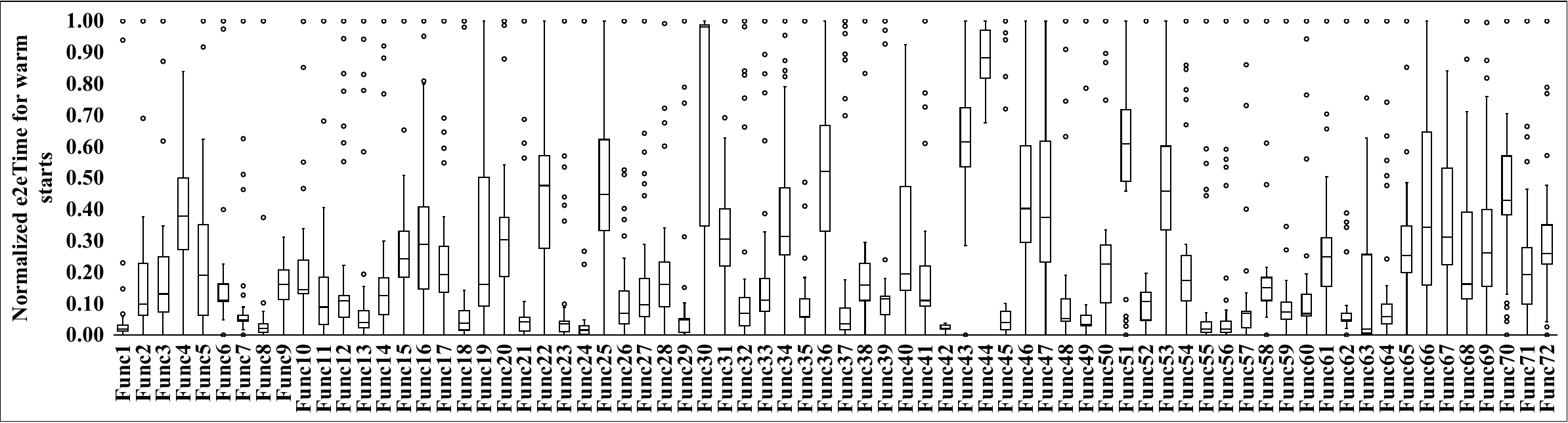}
    \caption{(RQ2) The normalized boxplot for the warm-start response latencies of 72 serverless functions.}
    \label{fig:boxplotwarm}
\end{figure*}

(2) \textit{Boxplot:} We use boxplots to visually show the distributions of response latencies of serverless functions. 
Before presenting the boxplots, we analyze the average latency of each serverless function. In cold starts, 36.11\% (26/72) of functions have an average latency of under 2 seconds, 65.28\% (47/72) have an average latency of under 5 seconds, and 83.33\% (60/72) are under 10 seconds. In warm starts, 36.11\% (26/72) have an average latency of under 1 second, 76.39\% (47/72) are under 5 seconds, and 83.33\% (60/72) are under 10 seconds. This highlights that most functions exhibit latencies in relatively small durations (e.g., milliseconds), in contrast to the minute- or hour-level latencies of traditional cloud applications.
Since serverless functions have different levels of latency granularity, we apply the commonly used min-max normalization method~\cite{normalization} to normalize each set of response latencies generated by the serverless function to the 0 to 1 interval. Fig.~\ref{fig:boxplotcold} 
and Fig.~\ref{fig:boxplotwarm} respectively show the normalized boxplot about the response latency of 72 serverless functions under cold start and warm start. 
The bottom of the box represents the value at the 25\textit{th} percentile, while its top is the 75\textit{th} percentile. The line in the box is the median, i.e., 50\textit{th} percentile. Dots outside of the whiskers are outliers. We observe that most response latencies (i.e., from the 25\textit{th} percentile to the 75\textit{th} percentile) of the serverless functions do not fall near the middle, i.e., 0.5 of the 0 to 1 interval. This occurs under both cold starts and warm starts. Moreover, the range sizes of most response latencies are inconsistent, indicating that there may not be a fixed distribution pattern for serverless function performance.

To further determine if the data distribution of serverless function performance is concentrated or skewed, we check whether the median of response latencies falls within an error interval. This interval represents the 1\% error above or below the middle value calculated from the maximum and minimum values. If the median is in this error interval, the data distribution for serverless function performance is determined to be concentrated, and otherwise, it is skewed. The results show that distributions of response latencies of more than 93\% of the serverless functions are skewed both under cold starts and warm starts. We also try to determine the direction in which most response latencies of the serverless function are biased to the boxplot. We calculate the sum of distances of all response latencies of each serverless function from its maximum and minimum values, respectively. The results show that most response latencies of over 93\% (cold starts: 69/72; warm starts: 67/72) of the serverless functions are skewed towards the corresponding minimum values, and another side of boxplots has a long tail. This can be observed visually in Fig.~\ref{fig:boxplotcold} and Fig.~\ref{fig:boxplotwarm}, where most boxplots are in the middle and lower part of the 0 to 1 interval.

\vspace{2.3mm}

\begin{mdframed}[linecolor=gray,roundcorner=12pt,backgroundcolor=gray!15,linewidth=3pt,innerleftmargin=2pt, leftmargin=0cm,rightmargin=0cm,topline=false,bottomline=false,rightline = false]
    \textbf{Finding 2}: The performance of serverless functions exhibits significant variance that cannot be ignored. Across multiple executions, the performance of serverless functions can vary by as much as 338.76\%, with an average variance of 42.28\%, indicating a large magnitude of the variance.
    Moreover, serverless function performance has different degrees of variance under cold and warm starts. 
    The coefficient of variance values for serverless function performance under cold starts are from 1.23\% to 31.18\%, while those of warm starts are 0.86\% to 42.81\%. These values differ by as much as 49 times, showing a wide range of variance. 
    Additionally, for 65.28\% of the serverless functions, the response latency variance during warm starts is more pronounced than that observed during cold starts.
\end{mdframed}

\section{RQ3: Reliability and Repetitions}\label{sec:credible}

In RQ3, we investigate the reliability of serverless function performance obtained at different repetitions. First, we use the response latencies of the serverless function at 50 repetitions as \textit{standard result} of serverless function performance, since most of the collected papers that report repetitions use no more than 50 times, as summarized in Section~\ref{sec:literature}. Then, we compare the serverless function performance obtained at \textbf{low repetitions} that are used in the collected papers and shown in Fig.~\ref{fig:PaperRepetition}, e.g., 3, 4, 5, 6, 10, and 20. 

\begin{figure}[t]
	\centering
    \includegraphics[width=0.7\textwidth]{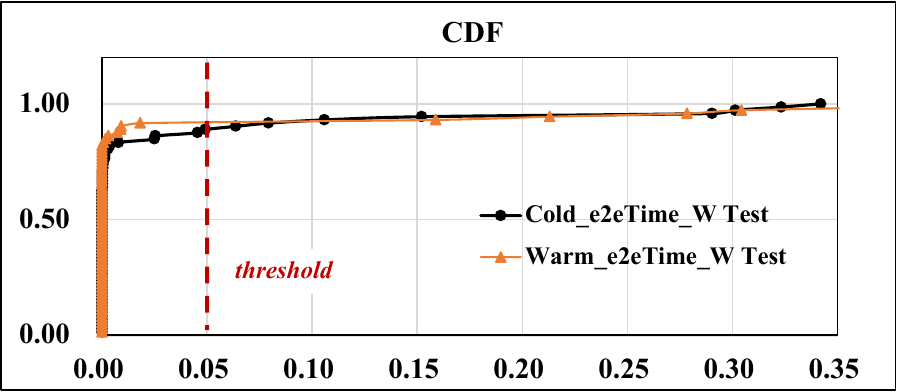}
    \caption{(RQ3) The CDF of $\rho$ values obtained from \textit{W} tests (normality checks), which are applied to the response latencies of each serverless function.}
    \label{fig:W}
\end{figure}

To facilitate performance analysis, we need to adopt appropriate statistical methods, either parametric or non-parametric, depending on the distribution of the performance results. For data that follows a normal distribution, parametric methods are suitable; for non-normal distributions, non-parametric methods are required. Thus, we first check whether serverless function performance follows a normal distribution.
We adopt the Shapiro-Wilk test~\cite{shapiro1965analysis} (abbreviated as \textit{W} test), which is considered the most powerful normality test in most situations~\cite{Wpopular}. In the \textit{W} test, its null hypothesis is that the performance results come from the population with the normal distribution. When running the \textit{W} test, we can obtain a $\rho$ value. At a $\rho$ value greater than 0.05, we can accept the null hypothesis to indicate that serverless function performance follows a normal distribution; otherwise, we reject the null hypothesis and describe the serverless function performance as following a non-normal distribution. We apply normality checks to the response latencies of 50 runs of each serverless function. Fig.~\ref{fig:W} shows the CDF about all $\rho$ values obtained from \textit{W} tests. We observe that most $\rho$ values are less than 0.05, i.e., rejecting the null hypothesis and presenting strong evidence for non-normality. Specifically, the response latencies of 88.89\% (64/72) and 91.67\% (66/72) of the functions follow a non-normal distribution in cold starts and warm starts, respectively. In this situation, non-parametric analysis methods, which do not assume normality, are more appropriate for the performance analysis of serverless functions, and they can also work for the normal distribution~\cite{maricq2018taming}.
Thus, we use the most common metrics of interest in non-parametric analysis, e.g., median, tail percentile, and their confidence intervals~\cite{maricq2018taming,uta2020big}, to compare performance. 

We calculate the median performance and tail performance (e.g., 90\textit{th} percentile) of the serverless function at different low repetitions. Meanwhile, we calculate the corresponding 95\% \textit{confidence interval for the median} and \textit{confidence interval for the 90\textit{th} percentile} at 50 repetitions, as adopted by the previous work~\cite{uta2020big}. The 95\% \textit{confidence interval for the median} or \textit{confidence interval for the 90\textit{th} percentile} represents the range in which we could find the true median or 90\textit{th} percentile with 95\% probability if we could perform infinite repetitions. Thus, when a median or 90\textit{th} percentile obtained at the low repetition lies outside the 95\% \textit{confidence interval for the median} or \textit{confidence interval for the 90\textit{th} percentile} obtained at the high repetition, it indicates that there is a 95\% probability that this median or 90\textit{th} percentile is inaccurate. The calculations of the \textit{confidence interval for the median} and \textit{confidence interval for the 90\textit{th} percentile} can refer to the work~\cite{le2010performance,maricq2018taming}.



\begin{figure}[t]
	\centering
    \includegraphics[width=0.7\textwidth]{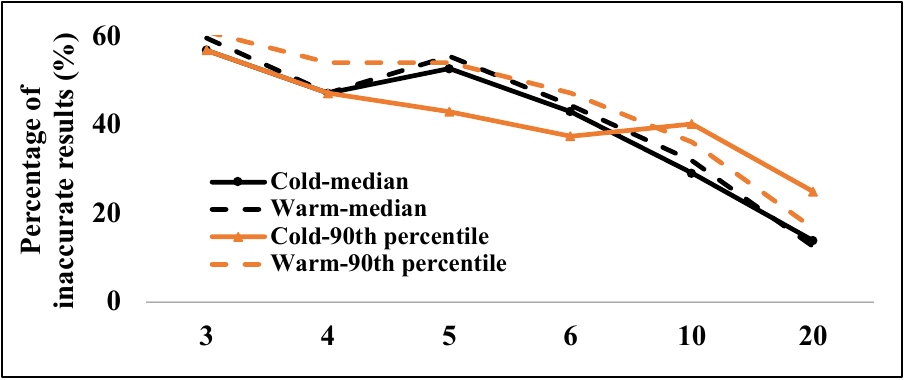}
    \caption{(RQ3) The percentage of serverless functions, whose medians and 90\textit{th} percentiles at low repetitions lie outside corresponding confidence intervals at 50 repetitions.}
    \label{fig:repetionsreliable}
\end{figure}

\textbf{Result:} We observe that the serverless functions produce inaccurate median performance and tail performance under low repetitions. Specifically, Fig.~\ref{fig:repetionsreliable} shows the percentage of the serverless functions in the case where the obtained median performance or 90\textit{th} percentile performance is inaccurate under different low repetitions. When the number of repetitions is 3, for 56.94\% (41/72) of the serverless functions in cold starts, the obtained median falls outside of the confidence intervals obtained at 50 repetitions, i.e., the obtained median performance is inaccurate, while warm starts have 59.72\% (43/72) of the serverless functions. For the 90\textit{th} percentile, 56.94\% (41/72) of the serverless functions have inaccurate tail performance at 3 repetitions (warm start: 61.11\% (44/72)). This indicates that experiments with low repetitions have a high risk of reporting unreliable performance results of serverless functions. We also observe that increasing the number of repetitions can make the percentage of inaccurate results decrease. However, at the 10-repetition most frequently used by the surveyed papers, 29.17\% (21/72) of the serverless functions still show inaccurate median performance, and 40.28\% (29/72) of serverless functions show inaccurate tail performance in cold starts. These results imply that serverless function performance is unreliable under low repetitions commonly used in most research papers, underscoring the significant consequences of neglecting the performance variance of serverless functions.

\begin{figure}[t]
	\centering
    \includegraphics[width=0.7\textwidth]{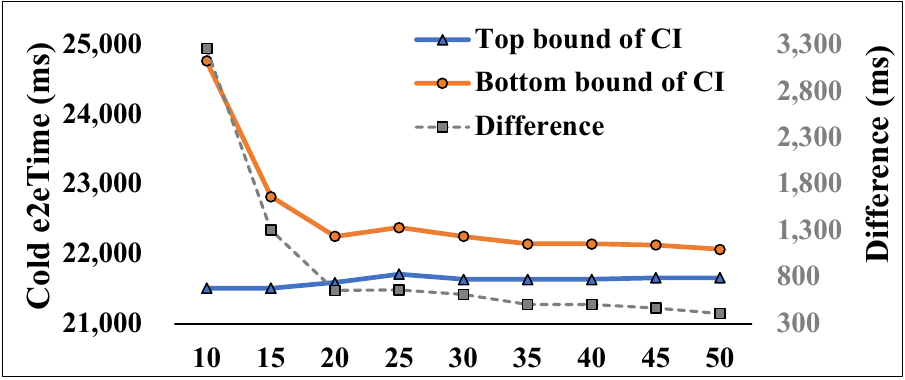}
    \caption{(RQ3) Changes in the top and bottom bounds of the \textit{confidence interval for the median} at different repetitions. \normalfont{An example is about \textit{Func2}.}}
    \label{fig:CIrepetition}
\end{figure}






We try to find regularities in how to determine the appropriate repetitions for serverless functions to obtain reliable performance. We first observe the changes in top and bottom bounds for the confidence interval obtained at different repetitions. As an example, Fig.~\ref{fig:CIrepetition} shows the changes in the bounds of the \textit{confidence intervals for the median}, with regard to \textit{Func2}. 
We do not plot the \textit{confidence intervals for the median} at smaller repetitions, e.g., 3, 4, 5, and 6 in Fig.~\ref{fig:repetionsreliable}, because these repetitions are also insufficient to calculate them~\cite{landis1977measurement,le2010performance}. Instead, we compare the number of repetitions from 10 to 50 in 5-step increments. From Fig.~\ref{fig:CIrepetition}, confidence intervals gradually become tight as the number of repetitions increases. The line representing the difference between the top and bottom bounds also shows a gradual downward trend. However, when the number of repetitions is small, e.g., from 20 to 25 repetitions, the bounds of the confidence interval may have slight fluctuations. It is reasonable that the distribution with a small number of performance data may not be stable. 
Overall, performing more repetitions for functions may achieve a tight \textit{confidence interval for the median}, which increases confidence in claiming the obtained results are close to the population distribution.

\begin{figure}[t]
	\centering
    \includegraphics[width=0.7\textwidth]{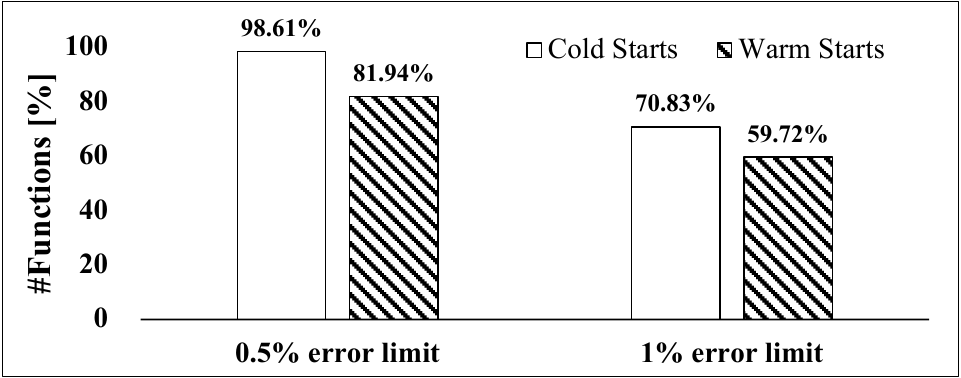}
    \caption{(RQ3) The percentage of serverless functions that take \textit{at least 50 times} to achieve a desired confidence interval about response latency, within 0.5\% and 1\% error demands.}
    \label{fig:determineRepetition}
\end{figure}

Leveraging the observation about tight confidence intervals for the median, we further explore how many repetitions may be required to achieve a sufficiently narrow confidence interval for serverless function performance. This interval is a desired interval representing satisfactory performance, where the empirical median differs from the observed true median by no more than the $r$ error margin at a given confidence level, e.g., 95\%. In other words, when the \textit{confidence interval for the median} obtained from results at a certain repetition drops within the $r$ error margin of the corresponding observed true median performance, the desired confidence interval is obtained to stop running the serverless function, and the current number of repetitions is regarded as the appropriate repetition for the serverless function. We show the results for $r$ = 0.5\% and 1\% in Fig.~\ref{fig:determineRepetition}. When $r$ = 0.5\%, in cold starts, 98.61\% (71/72) of the serverless functions require being executed at least 50 times to get the desired confidence interval, while warm starts have 81.94\% (59/72) of the serverless functions. After we relax $r$ to 1\%, there are still 70.83\% (51/72) serverless functions that need to be executed repeatedly at least 50 times in cold starts. There are 59.72\% (43/72) serverless functions in warm starts. However, according to the aforementioned Fig.~\ref{fig:PaperRepetition}, only 28.95\% of the research papers that report experiment repetitions execute serverless functions more than 50 times. Overall, these results indicate that serverless functions require far more repetitions than the ones commonly used in most surveyed papers.

\vspace{2.3mm}

\begin{mdframed}[linecolor=gray,roundcorner=12pt,backgroundcolor=gray!15,linewidth=3pt,innerleftmargin=2pt, leftmargin=0cm,rightmargin=0cm,topline=false,bottomline=false,rightline = false]
    \textbf{Finding 3}: 
    When a small number of repetitions, as commonly adopted in the collected papers, are executed, up to 61.11\% of serverless functions exhibit unreliable performance results.
    This underscores the significant consequences of neglecting the performance variance of serverless functions. 
    Additionally, up to 98.61\% of serverless functions require a minimum of 50 executions to achieve the desired performance, but only 28.95\% of the collected research papers, which report experiment repetitions, use repetition counts exceeding 50. These findings highlight the need for significantly more repetitions in serverless function experimentation than what is commonly employed in the existing literature.
    

    
\end{mdframed}

\section{Implications}\label{sec:guideline}

\textit{\textbf{Implications for researchers.}} \textbf{1) Reproducibility.} The research community has increasingly emphasized the importance of reproducibility and replicability in studies~\cite{anda2008variability,cavezza2014reproducibility,wwwDemirGUWHP22,rajiullah2019web,jueckstock2021towards,amiri2022modeling,yu2023brain,ardagna2018model}. Our study uncovers substantial performance variance in serverless functions, affecting research reproducibility and further deployment of cloud services. Insufficient documentation of performance measurement methods, including repetitions, hampers replication and result validation by other researchers. To enhance reproducibility in serverless computing, an open and explicit methodology is crucial. Researchers should clearly outline their performance measurement approach, including repetitions and rationale.  
\textbf{2)~Reliability.} Our study highlights concerns about research reliability in serverless computing due to insufficient consideration of performance variance. Inadequate repetitions and justification can diminish result reliability, risking erroneous conclusions. This can lead to misguided decisions and resource wastage in academia and industry. To address this, researchers can provide well-justified repetitions and comprehensive results, e.g., median, coefficient of variance, percentiles, confidence intervals, and distribution. Serverless functions exhibit two types of starts: cold and warm starts, with more pronounced performance variance in warm starts. Researchers can tailor experimental designs, including repetitions, for each start type to capture specific characteristics effectively. 
\textbf{3) Tailored performance testing.} Our evaluation of RQ3 shows that while 50 repetitions are commonly used in the serverless computing community, they are insufficient for achieving the desired performance for 98.61\% of serverless functions, due to significant performance variance, as shown in RQ2. In general, using more repetitions increases the accuracy and reliability of performance results by accounting for various potential impacts of the serverless platform. However, simply increasing repetitions for all serverless functions may not always be an effective solution because each function may require a different number of repetitions. This emphasizes the need for tailored performance testing in serverless computing, designed to ensure accurate and reliable performance measurement for serverless computing-based applications. Performance is a critical property of applications~\cite{savasci2023ddpc,zhang2021wisetrans,hounsel2020comparing,jayathilaka2017performance}, and performance testing has been a standard procedure for acquiring reliable performance of these applications~\cite{kbseHeLLLK021,he2021performance}. It involves iterative executions of the application-under-test with predefined inputs until a stopping criterion is met, thus mitigating the variance effect~\cite{sigsoftHeMS0PS19,kbseHeLLLK021,he2021performance}. Given the significant performance variance of serverless functions, there is research space to develop dedicated performance testing techniques and tailor the stopping criterion for this context. Furthermore, different from traditional cloud applications, serverless functions generally run for a short duration in milliseconds~\cite{wen2023FaaSlight,li2023serverless}. Thus, there is a need to use a finer performance accuracy analysis as the stopping criterion in serverless computing. Such approaches allow for efficient, targeted, and context-aware performance evaluation, regardless of when performance data for each function is captured or whether performance data experiences regular variances, thus avoiding the inefficiencies of simply increasing repetitions for all functions.


\noindent \textit{\textbf{Implications for software developers.}} \textbf{1) Mitigation strategies.} Given the substantial performance variance observed in serverless computing, it is essential for software developers to adopt strategies to mitigate this issue, especially when aiming for consistent user experiences. From our analysis of RQ2, the design of serverless function code may influence the magnitude of performance variance. Therefore, code optimization emerges as a potential mitigation strategy. This finding is consistent with various reports~\cite{bestpractice1,bestpractice2,bestpractice3} and existing studies~\cite{wen2023FaaSlight,taibi2022serverless}, which highlight the influence of coding practices on serverless function performance. For instance, optimizing the usage of cloud storage services is crucial, as access patterns to cloud storage can affect performance~\cite{bestpractice2}. Furthermore, removing unused dependencies from serverless functions can reduce unnecessary overhead, improving performance consistency~\cite{bestpractice2,wen2023FaaSlight}. These examples underscore the potential of adopting efficient coding practices to mitigate performance variance in serverless functions.
\textbf{2) Difference between start types.}  
In the development of serverless computing-based applications, software developers can carefully consider the distinct performance characteristics associated with both cold and warm starts. Our findings demonstrate that the response latency variance is more pronounced during warm starts than cold starts. To provide users with a seamless and consistent experience, software developers should ensure that these distinct characteristics do not impact the user experience under different start conditions.

\noindent \textit{\textbf{Implications for cloud providers.}} 
From our results, all the serverless platforms we study provide significant performance variance for serverless functions. Thus, cloud providers offering serverless environments should promise consistent quality of service when delivering the services by means of serverless computing. 
As highlighted in Finding 2, the performance of serverless functions can vary by as much as 338.76\%, with an average variance of 42.28\%. To ensure consistent quality of service, cloud providers could monitor performance changes of the serverless function as it is executed. If the magnitude of this change exceeds a specific threshold (e.g., average variance), cloud providers need to identify the function instance serving the current serverless function and conduct a detailed analysis of its resource usage to detect any abnormalities.

\section{Threats to validity}

\noindent \textbf{Selection of relevant papers.} Our empirical study involves analyzing research papers related to serverless computing. Given the difficulty of collecting all such papers, we select a representative set for analysis. This selection process may pose a threat to the validity of our results.
To mitigate this threat, we specifically gathered 110 research papers published in 77 top-tier conferences across different research communities. These top-tier conferences are known for including papers that have substantial impact and widespread recognition. The widespread recognition of these conferences underscores their status as trusted sources for analysis. This, in turn, enhances the representativeness of our data sources. However, we acknowledge that adding research papers from other journals or conferences would provide a more comprehensive analysis. In future work, we plan to expand more papers to further enhance the scope of our investigation.

\noindent \textbf{Manual examination of papers.} In RQ1, we manually label the three types of reporting information of each collected paper. This may pose a potential threat to the validity of our summarized results. To minimize this threat, the first two authors independently read the full text of the papers to determine specific information. Then we calculate the inter-rater agreement during the labeling process, and the obtained agreement values indicate a perfect agreement level and reliable labeling procedures. Additionally, to resolve conflicts, an experienced arbitrator, who has ten years of cloud computing experience, is involved in discussing and reaching an agreement.


\noindent \textbf{Root cause of performance variance.} In RQ2, we do not delve into the root cause of performance variance. The primary goal of our work is to raise awareness within the serverless computing community about the well-known performance variance problem in SE. Conducting a root cause analysis is beyond the scope of our study for several reasons. First, the serverless platforms commonly used by developers, such as AWS Lambda and Google Cloud Functions, are public and commercial services. While these platforms significantly reduce the management burden for developers, they also present challenges, as they operate as black boxes with opaque and uncontrollable policies. This lack of transparency makes it difficult to understand the underlying mechanisms that influence performance variance, complicating efforts to diagnose function performance issues. Second, the complexity of large-scale server management inherent to these serverless platforms further hinders the identification of specific root causes for performance variance. In future work, we plan to conduct a root cause analysis and connect the identified causes to the findings in literature, once serverless platforms provide more detailed information about their underlying runtime.

\noindent \textbf{Source of spatial variability.} Our study investigates the magnitude of performance variance in serverless functions across different platforms. In the current experiment, we focused on maintaining consistency in the service regions of the platforms used to ensure comparison fairness among different platforms. We acknowledge that spatial variability could be an important source influencing performance. In future work, we plan to extend our study to explore the impact of executing serverless functions across different service regions on performance variance.

\section{Related Work}\label{sec:relatedwork}


\textbf{Serverless computing.}
The research community has exhibited an increasing interest in serverless computing, as evidenced by the growing body of literature in this field~\cite{wen2022literature}. A systematic review of serverless computing research~\cite{wen2022literature} has been conducted, revealing a prominent focus on the performance of serverless computing in the existing literature.

Considerable research efforts have been dedicated to predicting and optimizing the performance of serverless computing. 
For performance prediction, previous studies~\cite{eismann2021sizeless,lin2020modeling} have utilized historical data, e.g., memory size or resource consumption, to predict serverless function performance. For performance optimization, 
Liu~et al.~\cite{wen2023FaaSlight} employed static program analysis techniques to optimize serverless function code and enhance its performance. 
Qi~et al.~\cite{qi2022spright} presented a shared-memory framework and evaluated its performance improvement effectiveness on various serverless functions. 
For these studies, ignoring the performance variance problem can lead to inaccurate performance predictions and suboptimal optimization outcomes.

Additionally, there have been empirical studies that focus on characterizing the performance of serverless computing. For example, 
Wang~et al.~\cite{WangATC2018-63} characterized serverless platform performance, examining scalability, cold start, etc. While they acknowledged the variability in latency during instance preparation of cold starts across multiple runs, they did not systematically investigate performance variance, as it was not the primary focus of their study. Wen~et al.~\cite{wen2021characterizing} performed a measurement study to evaluate the performance of commodity serverless platforms using different serverless functions.
Similarly, Eismann~et al.~\cite{eismann2022case} explored the stability of performance measurements on serverless platforms, specifically investigating various load or concurrency configurations. Different from these empirical studies, our objective is to examine the awareness of researchers about the performance variance problem and characterize the magnitude of this variance in serverless computing.

\noindent \textbf{Performance variance.} 
Performance variance is a well-known problem in SE, attracting extensive research efforts for analysis. 
As cloud computing offers efficient resource management, it has gained widespread adoption. 
For instance, 
He~et al.~\cite{he2019statistics,he2021performance} executed cloud applications hosted in a traditional cloud computing paradigm, Infrastructure-as-a-Service~\cite{bhardwaj2010cloud}. They found that it is challenging to obtain accurate performance. 
Laaber~et al.~\cite{laaber2019software} studied the impact of cloud environments on result variability.
For other systems, 
Pham~et al.~\cite{pham2020problems} quantified the variance of deep learning systems regarding models' accuracy, varying by up to 10.8\%. 
Qian~et al.~\cite{qian2021my} explored the variance of fairness metrics in deep learning systems, identifying significant variance of up to 12.6\%.  
Georges~et al.~\cite{georges2007statistically} delved into the performance variance of Java systems and highlighted the importance of statistically rigorous data analysis.


Although the significance of performance variance has been acknowledged, current serverless computing research still frequently overlooks this well-known problem.
Moreover, the magnitude of performance variance in the new programming model - serverless functions, remains unclear. While it is true that many practitioners and researchers may have been apprehensive about the performance variance of serverless computing, there has not been solid scientific evidence of the magnitude of such an issue thus far. In this paper, we provide a comprehensive study to shed light on the oversight of performance variance in previous serverless computing research work and characterize the magnitude of this variance, emphasizing the significant consequences of disregarding this crucial aspect.
Without this, all we are left with is a ``belief'' rather than quantitative scientific evidence.




\section{Conclusion}\label{sec:conclusion}

We conducted an empirical study to highlight the lack of awareness in serverless computing research regarding the well-known performance variance problem in software engineering. To this end, we first collected and analyzed 99 research papers related to serverless computing performance, showing that attention to this performance variance is still in the infancy stage. Then, we conducted a measurement study to illustrate the substantial magnitude of performance variance in serverless computing. Specifically, 
we analyzed the end-to-end response latencies of 72 serverless functions collected from these papers obtained over multiple runs. We observed a significant performance variance, with a maximum variance of 338.76\% (44.28\% on average) among different runs. We further analyzed the reliability of serverless function performance obtained by executing a low number of repetitions, as commonly done in our collected papers. We found that 61.11\% of the serverless functions have unreliable performance. This underscores the significant consequences of neglecting this critical aspect of performance variance.
Moreover, 98.61\% of the serverless functions required being executed at least 50 times to achieve reliable performance, but only 28.95\% of the collected papers that report experiment repetitions execute serverless functions more than 50 times. This implies that serverless functions require far more repetitions than the ones commonly used in the literature.

\section{Data Availability Statements}


The detailed information for collected research papers and serverless functions are available in a public GitHub repository~\cite{ourdata}. Moreover, we provide the deployment package and raw performance data of each serverless function, as well as code scripts used in our study.

\section*{Declarations}


\subsection*{Funding and/or Conflicts of interests/Competing interests}

This work is supported by the National Natural Science Foundation of China under Grant No. 62032003.  
\textbf{The authors declare that they have no conflict of interests and competing interests.}






\bibliographystyle{spmpsci}      
\bibliography{template}   


\end{document}